\RequirePackage{fix-cm}
\documentclass[twocolumn,epjc3]{svjour3}  
\smartqed  
\RequirePackage{graphicx}
\RequirePackage{amsfonts}
\RequirePackage{amssymb}
\RequirePackage{amsmath}
\RequirePackage{infwarerr}
\RequirePackage{ltxcmds}
\RequirePackage{epstopdf}
\RequirePackage{bm}
\RequirePackage{color}
\RequirePackage[T1]{fontenc}
\RequirePackage[utf8]{inputenc}
\RequirePackage{widetext}
\RequirePackage{cuted}
\RequirePackage{ctable}
\RequirePackage{subfig}

\newcommand{\beq}{\begin{equation}}
\newcommand{\eeq}{\end{equation}}
\journalname{Eur. Phys. J. C}
\begin{document}

\title{Thermodynamical and dynamical equilibrium of a self-gravitating uncharged thin shell 
}


\author{Santiago Esteban Perez Bergliaffa\thanksref{e1,addr1}
        \and
        Marcelo Chiapparini\thanksref{e2,addr1}
        \and
        Luz Marina Reyes\thanksref{e3,addr2} 
}

\thankstext{e1}{e-mail: sepbergliaffa@gmail.com}
\thankstext{e2}{e-mail: chiapparini.uerj@gmail.com}
\thankstext{e3}{e-mail: luzmarinareyes@gmail.com}

\institute{Departamento de F\'isica Te\'orica, Instituto de F\'isica, Universidade do Estado do Rio de Janeiro,\\
Rua S\~ao Francisco Xavier 524, Maracan\~a, CEP 20550-900, Rio de Janeiro, Brasil. \label{addr1}
           \and
           Departamento de Ciencias Computacionales, CUCEI, Universidad de Guadalajara. Av. Revoluci\'on 1500, 44430, Guadalajara 
 Jal., M\'exico. \label{addr2}
}

\date{Received: date / Accepted: date}

\maketitle

\begin{abstract}
The dynamical stability of  massive thin shells with a given equation of state (EOS) (both for the barotropic and non-barotropic case)
is here compared with the results coming from thermodynamical stability.
Our results show that the restrictions in the parameter space of equilibrium configurations of the shell following from thermodynamical stability are much more stringent that those obtained from dynamical stability. As a byproduct, we furnish evidence that the link between 
the 
maximum mass along a sequence of equilibrium configurations and the onset of dynamical stability is valid for EOS of the type $P=P(\sigma, R)$.
\keywords{Dynamical stability \and Thermodynamical stability \and Massive thin shells \and Equation of state}
\end{abstract}

\section{Introduction}
\label{intro}
Self-gravitating thin shells are solutions of a given gravitational theory describing two regions separated by an infinitesimally thin region where matter is confined. Such a system  conjugates the notions of vacuum,  typical of black holes, together with the presence of matter, which may be described via statistical mechanics and thermodynamics.
Thin shells have been frequently employed to probe thermodynamical properties  of black holes, see for instance  \cite{York1986,Lemos2015,Lemos2015b,Lemos2016,Lemos2017} and, since they
can be taken to their own gravitational
radius, they can be transformed into quasi-black holes
\cite{Lemos2007}, and used to calculate black hole properties (see for instance \cite{Lemos_2011})
\footnote{For a complete list of applications of the thin-sell formalism see
\cite{Kijowski2006}.
}.
In view of these applications, it is important to determine whether relevant thin-shell configurations are stable, both thermodynamically and dynamically.
The thermodynamic stability of a spherically symmetric thin shell in which the interior region is Minkowski's spacetime and the exterior given by Schwarzschild's geometry was studied in \cite{Martinez1996}, while the linear dynamical stability of such systems under radial perturbations was analyzed in
\cite{Brady1991} for a linear EOS, and in
\cite{Habib2017}
for a general EOS. Since these two types of stability yield inequivalent restrictions on the parameter space \cite{Martinez1996}, 
we shall present here 
the results of imposing both types of stability on a neutral thin shell configuration, for different barotropic
equations of state, and also for EOS of the type $P=P(\sigma,R)$. The latter
have been used in various settings such as wormholes 
(see for instance  \cite{Rahaman2007,Varela2013,Garcia2011}), stars in Eddington-inspired Born-Infeld gravity 
 \cite{Kim2014} and gravastars \cite{MartinMoruno2011}, and in cosmology 
\cite{Guo2005,Debnath2007}, among others.
For a given EOS, we shall  determine the region of the parameter space of equilibrium configurations of the shell compatible with both types of stability, and with the dominant energy condition (DEC).
As a byproduct, we shall obtain evidence 
supporting the extension 
of the results linking the maximum mass 
with the onset of instability \cite{LeMaitre2019} to EOS of the type $P=P(\sigma,R)$.

 The paper is organized as follows. In Section \ref{junction} we shall present a brief review of the relevant equations for the thin shell in equilibrium (mostly following \cite{Lemos2015}).
The equations obeyed by the perturbed shell for an EOS of the type $P=P(\sigma,R)$ will be introduced in Section \ref{dynamical}. In Section \ref{mR} we shall present 
the analysis of the dynamical stability of the shells for different barotropic EOS, along with the corresponding
$m( R)$ diagrams. Section \ref{sec:5} is devoted to a non-barotropic EOS. The restrictions following from thermodynamical stability 
are exhibited in Section  \ref{sec:6}. 
In Section  \ref{sec:7} we determine the set of equilibrium states of the shell that are both dynamically and thermodynamically stable. 
Our closing remarks are presented in Section  \ref{sec:8}.
\section{Junction conditions and properties of the thin-shell}
\label{junction}
Let us consider a two-dimensional timelike massive
shell $\Sigma$ with radius $R$. The shell divides the spacetime in two parts: i) an inner region $r<R$, with flat geometry, and ii) an outer region $r>R$, in which the geometry is described by 
the Schwarzschild line element. In this way, we can express the metric in both regions as follows:
\begin{equation}\label{RN1}
 ds_{I}^2 =
 -f_{I}{dt_{I}}^2+g_{I}dr^2 + r^2d\Omega ^2.
\end{equation}
Here $I = o, i$ refers either to the outer or inner region, and the functions $f_{I}$
 and $g_{I}$ are given by
\begin{equation}\label{RN2}
 f_{i}=g_{i}=1,\quad f_{o}={1-\frac{2m}{r}
 },\quad g_{o}=\left(1-\frac{2m}{r}\right)^{-1},
\end{equation}
\noindent where $m$ is  the ADM mass,  $d\Omega ^2\equiv d\theta ^2+\sin{\theta}^2d\phi^2$. \\
The metric $h_{ab}$, defined  on $\Sigma$, \emph{i.e.} for $r=R$,  is that of a 2-sphere, and 
can be written as
\begin{equation}\label{RN9}
 ds_{\Sigma}^2 =h_{ab}dy^a dy^b=-d\tau^2 + R^2(\tau)d\Omega ^2,
\end{equation}
where $y^a=(\tau,\theta,\phi)$ and $R$ is a function of $\tau$, the proper time for an observer located on the shell, in the dynamical case. The application
of the thin-shell formalism developed in \cite{Israel1966} to 
join the two spacetimes specified in Eqs. (\ref{RN1}) and 
(\ref{RN2})
requires 
the induced metric $h_{ab}$ to be continuous on the shell, and the
discontinuity in the extrinsic curvature to be proportional to the stress-energy tensor on the shell, denoted by ${\cal S^{\mu}_{\:\nu}}$. The latter is given by the surface energy density $\sigma$ and the tangential pressure $p$ which, for a static shell, are as follows (see for instance
 \cite{Martinez1996}):
\begin{align}
 {{\cal S}^\tau}_\tau &\equiv  \sigma_0 =\dfrac{1-\sqrt{1-\frac{2m_0}{R_0}}}{4\pi R_0},\label{RN3}\\
{{\cal S}^\theta}_\theta = {{\cal S}^\phi}_\phi &\equiv P_0 =\frac{\sqrt{1-\frac{2m_0}{R_0}}-1}{8\pi R_0}+\frac {m_0}{8\pi R_0^2 \sqrt{1-\frac{2m_0}{R_0}}}\label{RN4},
\end{align}
where the subindex 0 means that the quantities are evaluated at the equilibrium configuration.\\
The proper mass of the shell, denoted by $M$, is given by $M=4\pi R_0^2\sigma_0$. 
The junction conditions also imply that 
the ADM mass is given by
\begin{equation}\label{RN7}
m_0=M-\frac{M^2}{2R_0}.
\end{equation}
Hence we can write
\begin{align}
P_0(M,R_0) &= \frac{M^2}{16\pi R_0^2(R_0-M)},\label{pMR} \\
\sigma_0(M,R_0) &= \frac{M}{4\pi R_0^2}. \label{sigmaMR}
\end{align}
We shall assume that $\sigma_0$ and $P_0$ are non-negative (hence $M>0$) and $R_0\geq 2m_0$ \footnote{We use units such that $[M]=[m]=[R]=L$ and $[P]=[\sigma]=1/L$.}.
\\
Notice that by inverting
Eqs. (\ref{pMR}) and (\ref{sigmaMR}) 
we obtain $R_0(P_0,\sigma_0)$ and $M(P_0,\sigma_0)$ 
respectively given by
\begin{align}
    R_0(P_0,\sigma_0)&=\frac{P_0}{\pi\sigma_0(\sigma_0+4P_0)}, \label{rps}\\
    M(P_0,\sigma_0)&= \frac{4P_0^2}{\pi\sigma_0(\sigma_0+4P_0)^2}. \label{mps}
\end{align}
Using Eqs. (\ref{RN7}) and (\ref{mps}) it follows that
\begin{align}
    m_0(M,R_0)&= 4\pi\sigma_0 R_0^2(1-2\pi\sigma_0 R_0).
\end{align}
The final equations for the mechanical equilibrium of the shell are then
\begin{align}
    R_0(P_0,\sigma_0)&=\frac{P_0}{\pi\sigma_0(\sigma_0+4P_0)}, \label{rps2}\\
    m_0(P_0,\sigma_0)&=4\pi\sigma_0 R_0^2 (1-2\pi\sigma_0 R_0). \label{mps2}
\end{align}
These equations will be used in the analysis of the linear stability and
to build
the $m_0=m_0(R_0)$ diagrams for a given EOS, as we shall see in Sect.\ref{mR}.
\\
Before moving to the dynamical stability of the shell, let us introduce the redshift
of the shell, defined as
\begin{equation}\label{RN5}
k=\sqrt{1-\frac{2m_0}{R_0}}.
\end{equation}
It follows that 
$M=R_0(1-k)$.

\section{Dynamical stability}
\label{dynamical}
We shall outline here the steps that lead to the condition for the linear dynamical stability of the shells introduced in the previous section.
While Eqs. (\ref{pMR}) and (\ref{sigmaMR}) describe the equilibrium state of the shell, the corresponding expressions for a dynamical shell are
(see for instance \cite{Habib2017,Garcia2011}).   
\begin{align}
\sigma=&\frac{1}{4\pi}\frac{\sqrt{f_o(R)+\dot R^2}-\sqrt{f_i(R)+\dot R^2}}{R},
\label{sigmat}\\
P=&\frac{1}{8\pi}\left[ \frac{2\ddot R+f_o'(R)}{2\sqrt{f_o(R)+\dot R^2}}-\frac{2\ddot R+f_i'(R)}{2\sqrt{f_o(R)+\dot R^2}}\right.\nonumber\\
&+\left.\frac{\sqrt{f_o(R)+\dot R^2}-\sqrt{f_i(R)+\dot R^2}}{R}
\right],
\end{align}
where $f_i(R)$ and $f_o(R)$ are defined in Eq. \eqref{RN2}, and the overdot denotes the  derivative with respect to $\tau$. These quantities obey the equation that follows from the conservation of ${\cal S}^\mu_{\;\nu}$, namely
\begin{equation}
    \frac{d\sigma}{dR}+\frac{2}{R}(P+\sigma) = 0.
    \label{cons}
\end{equation}
A radial perturbation of an equilibrium configuration with $R=R_0$ causes $R$, $\sigma$ and $P$ to become functions  of $\tau$. Assuming 
an EOS of the type 
$P=P(\sigma, R)$, it follows from 
Eqs. (\ref{sigmat})
and (\ref{cons}) that 
the evolution of the shell 
is governed by the equation  \cite{Habib2017}
\begin{equation}
    \dot R + V(R,\sigma(R))=0,
\end{equation}
where 
{\small
\begin{align}
V (R)=& \frac 1 2 \,({ f_i} \left( R \right) +{f_o} \left( R \right)) -{
\frac {1}{64}}\,{\frac { \left( {f_i} \left( R \right) -{f_o}
 \left( R \right)  \right) ^{2}}{{\pi}^{2}{R}^{2} \sigma
 \left( R \right)   ^{2}}}\nonumber\\&-4\,{\pi}^{2}{R}^{2}  \sigma
 \left( R \right)  ^{2}.
\end{align}
}
The linear stability of the shell can be studied by expanding
the potential $V(R)$
around the equilibrium state
up to second order in $x=R-R_0$, hence obtaining
\begin{align*}
\frac{d^2x}{d\tau^2}+\omega_0^2 x = 0.
\end{align*}
Stability implies that
\begin{align}
\omega_0^2\equiv\left.\frac{d^2V}{dR^2}\right|_{R_0}>0.
\end{align}
The calculation of $\omega_0^2$ involves $\frac{d\sigma}{dR}$ (which is given by Eq. (\ref{cons})), and $\frac{d^2\sigma}{dR^2}$, which is obtained by taking the derivative of Eq. (\ref{cons}). For a general EOS
of the type 
$P=P(R,\sigma)$, the $\omega_0^2$ is given by
 \cite{Habib2017}:
 \begin{widetext} 
\begin{eqnarray}
\omega_0^2 &=& 
-16\pi\,{\frac {H_0F_0 }{F_0-H_0}}\,{\Omega_{10}}+2
\,\frac {H_0(2F_0^2-f'_{10}R_0)-F_0(2H_0^2-f'_{20}R_0)}{(F_0-H_0)R_0^2}\Omega_{20}+
\\ \nonumber
&& \frac{\left[ 4F_0^4-2R_0F_0^2(f'_{10}+R_0f''_{10})+R_0^2f_{10}^{'2}\right]H_0^3-\left[4H_0^2-2R_0H_0^2(f'_{20}+R_0f''_{20})+R_0^2f_{20}^{'2} \right]F_0^3}{2(F_0-H_0)F_0^2H_0^2R_0^2},
\end{eqnarray}
\end{widetext}
\noindent where $F_0\equiv\sqrt{f_{i0}}$, $H_0\equiv\sqrt{f_{o0}}$, and
\begin{align}
\Omega_{10}&\equiv\left.\frac{dP}{dR}\right|_{R_0} \label{defo1},\\
\Omega_{20}&\equiv\left.\frac{dP}{d\sigma}\right|_{R_0} \label{defo2} .
\end{align}
The line dividing stability from instability is given by $\omega_0^2=0$ which,
using 
Eq. (\ref{RN2}),
leads to the following expression for the critical values of
$\Omega_{20}$:
\begin{align}
\Omega_{20c} &=  \frac{A\;\Omega_{10c}+B}
{D}
\label{omega2nb}
\end{align}
\noindent where
\begin{widetext} 
\begin{eqnarray*}
A&=&
\pi{R_0}^{3}
\left[
8\left( 2 \,{R_0}^{2}+6 {m_0}^{2}-7 \,{R_0}m_0
 \right) 
 u
 +
 72 \,{R_0}^2 m_0+32
\,{m_0}^{3}-96 \,{R_0}{m_0}^{2}-16 \,{R_0}^{3}
\right], \\
B&=&{m_0}^{2}( 3\,{m_0}
-15\,R_0)+{R_0}^{2}
(14\,m_0-4\,{R_0}) 
u+
R_0 [
m_0 (
27\,{m_0}{R_0}^{2}-13\,{m_0}^{2}-
18\,{R_0}^{2}) +4\,{R_0}^{3}],\\
D&=&2\left[ {R_0}^{2}(
4\,{R_0}-18\,m_0)+
{m_0}^{2}
(23\,R_0-6\,{m_0}) \right] 
u+  
R_0 [
-4\,{R_0}^{3}+
m_0(
22\,{R_0}^{2}-39\,{m_0}{R_0}+22\,{m_0}^{2})],
\end{eqnarray*}
\end{widetext} with $u\equiv\sqrt{R_0(R_0-2m_0)}$.
Eq. (\ref{omega2nb}), valid for an arbitrary EOS of the form $P=P(\sigma, R)$,
determines regions of stability in a certain space of parameters. In particular, in the barotropic case, $\Omega_{10}=0$, and 
$\Omega_{20c}=B/D$
 defines the surface $\Omega_{20c} = \Omega_{20c} (m_0, R_0)$. Any equilibrium configuration with $(m_0,R_0)$ such that $\Omega_{20} $ is greater that $\Omega_{20c}$ will be stable \footnote{
Equation (\ref{omega2nb})
was used in \cite{Habib2017} to analyze the stability of two systems (a thin shell connecting two spacetimes of cloud of strings, and 
a thin shell connecting vacuum to Schwarzschild)
in the 
$(\Omega_{10c},\Omega_{20c}) $ plane without specifying the EOS.}. 
In the non-barotropic case, Eq. \eqref{omega2nb}
defines a 3-d surface by 
$\Omega_{20c} = \Omega_{20c} (m_0, R_0, \Omega_{10c})$.\\
When a specific EOS is chosen, there are  other constraints that must be taken into account.
As we shall see in Section
\ref{mR}, 
using 
the equilibrium equations
(\ref{RN3}) and (\ref{RN4})
for a given EOS, we can obtain the derivatives of the EOS as $\Omega_{10}=\Omega_{10}(m_0,R_0;\kappa)$, and $ \Omega_{20} = \Omega_{20}(m_0,R_0;\kappa)$, 
where $\kappa$ denotes the parameters of the EOS.
The EOS and the equilibrium equations also yield $m=m(R_0;\kappa)$. Combining the
latter with the equation \eqref{defo2} for $\Omega_{20}$
we obtain
$\Omega_{20}=\Omega_{20}(R'_0)$,  
where $R_0'$ is the radius of the equilibrium configuration normalized using the dimension-full parameter of the EOS. Using 
$m=m(R_0;\kappa)$
and the equation for $\Omega_{10}$
in 
the equation \eqref{omega2nb} for $\Omega_{20c}$, we
obtain
$\Omega_{20c}=\Omega_{20c}(R'_0)$,
The dynamically stable configurations for the given EOS will be those with $\Omega_{20}(R'_0) > \Omega_{20c}(R'_0)$.
We shall consider in the next section the
dynamical stability 
of the shell presented in Section \ref{junction}
for several relevant examples of EOS. 

\section{Dynamical stability for different EOS and the $m_0=m_0(R_0)$ curve.}
\label{mR}
Let us apply next the discussion of the previous section to several EOS of interest, to determine the regions of dynamical stability in the $m_0\times R_0$ plane, as well as the actual equilibrium states given by the $m_0=m_0( R_0)$ curve. It is important to note that the results  from the dynamical stability analysis we shall present
are in agreement with 
the criterion of the maximum of the 
$m_0=m_0(R_0)$ diagram, a fact that was proved in \cite{LeMaitre2019} for the case of a barotropic EOS. Our results suggest that the criterion is also valid for EOS of the form $P=P(\sigma, R)$.
All the $m_0=m_0( R_0)$ curves stop at the point where the DEC $P\leq\sigma$ ceases to be satisfied.

\subsection{Quadratic and barotropic EOS}
\label{quad}
\begin{figure}
   \centering
   \includegraphics[width=0.5\textwidth]{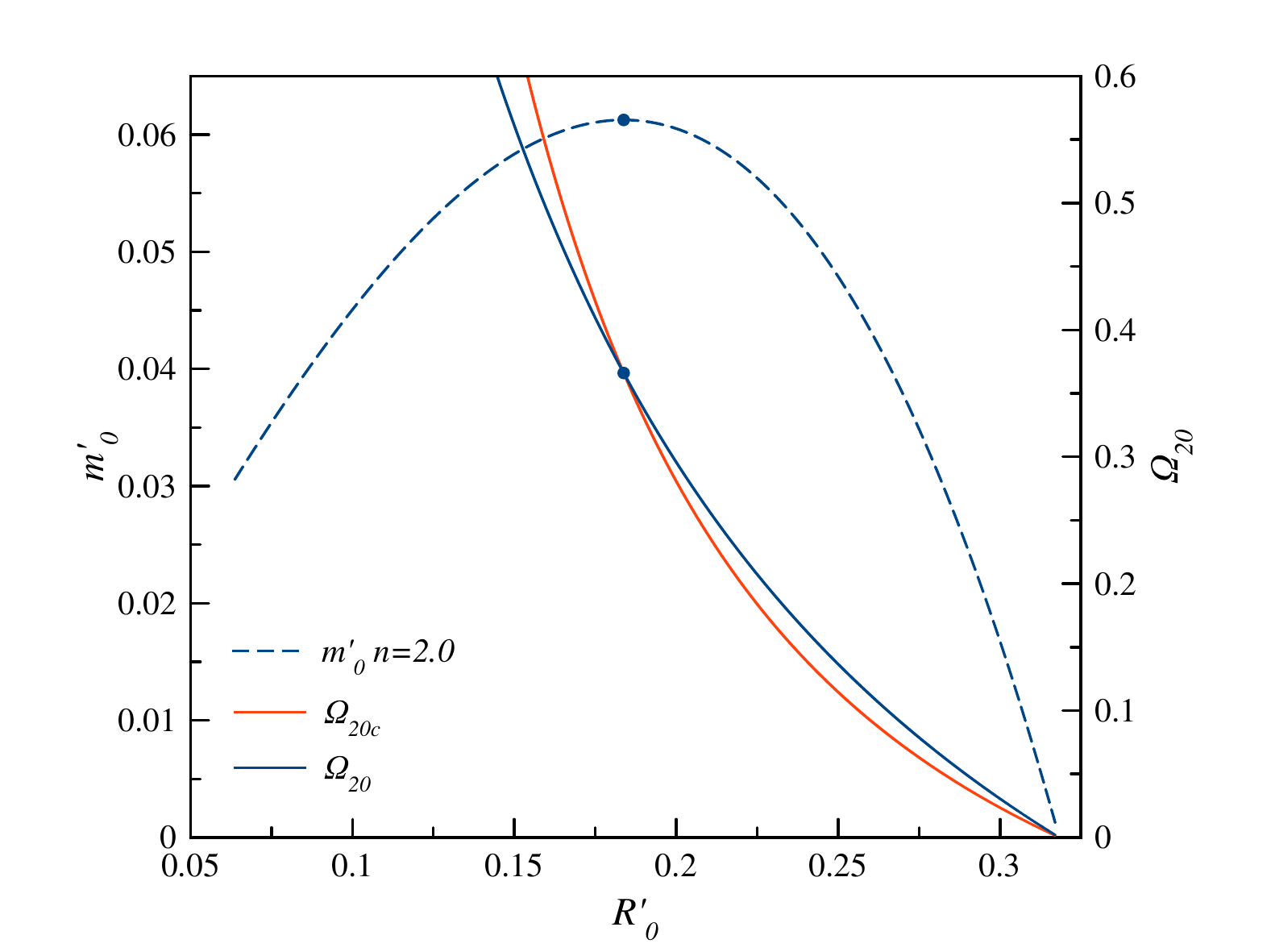}
\caption{The figure shows the $m'_0(R'_0)$ diagram for the EOS $P=\beta\sigma^2$, as well as the critical values of $\Omega_2$ that follow from dynamical stability (see Eq. \eqref{omega2nb}), and those that follow from the EOS, in red and blue, respectively.}
\label{fig1}
\end{figure}

We shall start with the example of a barotropic EOS, given by
\beq
P = \beta \sigma ^2,
\label{eosb2}
\eeq
where $[\beta]=L$. 
Such an equation models the non-relati-vistic limit of a two-dimensional ideal Fermi gas at $T=0$, discussed in Subsection \ref{r2db}. 
The constant $\beta$ 
can be used to render 
dimensionless all the variables in the problem as follows: $\sigma'=\beta\sigma$, $P'=\beta P$, $R'=R/\beta$, and $m'=m/\beta$. 
In these variables, Eq. (\ref{eosb2}) reads $P'=\sigma'^2$. 
Substituting $P'$ 
into Eq. (\ref{rps2}) we obtain $\sigma_0=\sigma_0(R_0)$
which, in dimensionless form,
is given by
\beq
    \sigma'_0(R'_0)=\frac{1}{4}\left(\frac{1}{\pi R'_0}-1\right). \label{slb2}
\eeq
Using $\sigma'_0(R'_0)$ in Eq. (\ref{mps2}) we obtain the 
corresponding
$m_0'(R_0')$ relation:
\beq
    m'_0(R'_0)=\frac{1}{2}R'_0\left(1-\pi^2 R_0^{'2}\right). \label{mlrlnr}
\eeq
In the domain $R'_0\ge 0$, the function $m_0'(R_0')$ 
has a maximum for $R'_{0max}=1/\left(\pi\sqrt{3}\right)$, resulting in 
$m'_{0max}=1/\left(3\pi\sqrt{3}\right)$.
The $m'_0=m'_0(R'_0)$ curve for this case is shown in Fig. \ref{fig1}. 
In the low energy density limit $(\sigma'\rightarrow 0)$, we can see that $m'\rightarrow 0$ and $R'\rightarrow 1/\pi$ (thus indicating that this EOS allows shells with very low mass). In the high energy density limit $(\sigma'\rightarrow \infty)$, we have $m'\rightarrow 0$ and $R'\rightarrow 0$, however this region of the curve is not depicted because it violates the DEC. Notice also that $m_0(R'_0)$ given by (\ref{mlrlnr}) satisfies $2m'_0<R'_0$, which is a constraint from the theory of thin shells \cite{Lemos2015}.\\
Next we indicate how to obtain the 
curves for $\Omega_{20c}$ and $\Omega_{20}$ of Section \ref{dynamical}.
Using Eq. (\ref{mlrlnr}) in Eq. (\ref{omega2nb}) with $\Omega_{10c}=0$, the curve $\Omega_{20c}=\Omega_{20c}(R'_0)$ follows, plotted in red in Fig. \ref{fig1}.
For $\Omega_{20}$ we use that $\Omega_{20}\equiv
\left.\frac{dP}{d\sigma}\right|_0= 2\sigma'_0$, where 
$\sigma'_0$ is given by Eq. (\ref{slb2}). This yields $\Omega_{20}=\Omega_{20}(R'_0)$, plotted in blue in Fig. \ref{fig1}.
The curve $\Omega_{20}(R'_0)$ intersects $\Omega_{20c}(R'_0)$
exactly at the value of $R'_0$ corresponding to the maximum of the $m'_0(R'_0)$ curve, and all the configurations
on the curve to the right of this point ($\Omega_{20}>\Omega_{20c}$) are stable.

\subsection{A relativistic EOS (EOS I)} \label{r2db}
\begin{figure}
    \centering
    \includegraphics[width=0.5\textwidth]{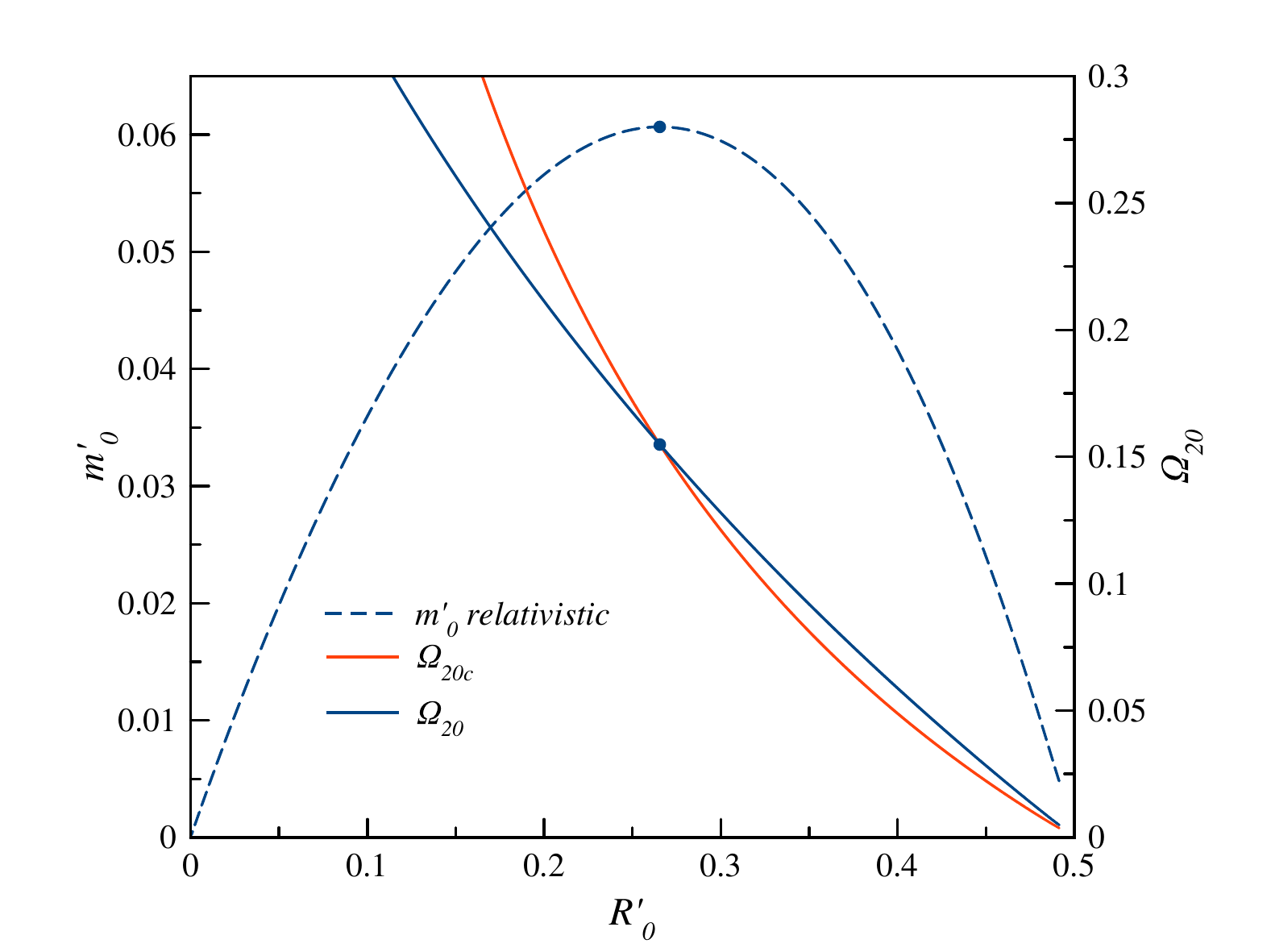}
    \caption{
        The figure shows the $m'_0=m'_0(R'_0)$ diagram for the relativistic EOS defined by Eqs. \eqref{dpldsl0} and \eqref{dpldsl02}, as well as the critical values of $\Omega_2$ that follow from dynamical stability  
        and those that follow from the EOS, in red and blue, respectively.
        }
    \label{rel}
\end{figure}

As shown in the Appendix, 
the EOS for a system of non-interacting relativistic fermions in 2d at $T=0$ is given by
\begin{align}
    \sigma_0&= \frac{\alpha}{3\pi}\left[\left(x^2+1\right)^{3/2}-1\right], \label{dpldsl0} \\
    P_0&=\frac{\alpha}{3\pi}\left[\frac{1}{2}(x^2-2)\sqrt{x^2+1}+1\right], \label{dpldsl02}
\end{align}
with
$x=p_F/mc$, $\alpha=mc^2/\lambda^2$ and $\lambda=\hbar/mc$. Following the same steps as in Subsection \ref{quad}, where the dimensionless variables are now $\sigma'=\sigma/\alpha$, $P'=P/\alpha $, $R'=\alpha R$, and $m'=\alpha m$, we obtain
the plots shown in 
Fig. \ref{rel}.
The values to the right of the maximum at 
 $(R'_{0max},m'_{0max})= (0.26545,0.060666)$ are stable. In the low energy density limit $(\sigma'\rightarrow 0)$, the plots show that $m'\rightarrow 0$ and $R'\rightarrow 0.5$
(indicating that this EOS also allows shells with very low mass), while in the high energy density limit $(\sigma'\rightarrow \infty)$,  $m'\rightarrow 0$ and $R'\rightarrow 0$. The DEC is satisfied along the whole curve as expected together with the constraint 
$2m_0'<R_0'$.

\subsection{A more general barotropic EOS (EOS II)} \label{beos2}
\begin{figure}
  \centering
 \includegraphics[width=0.5\textwidth]{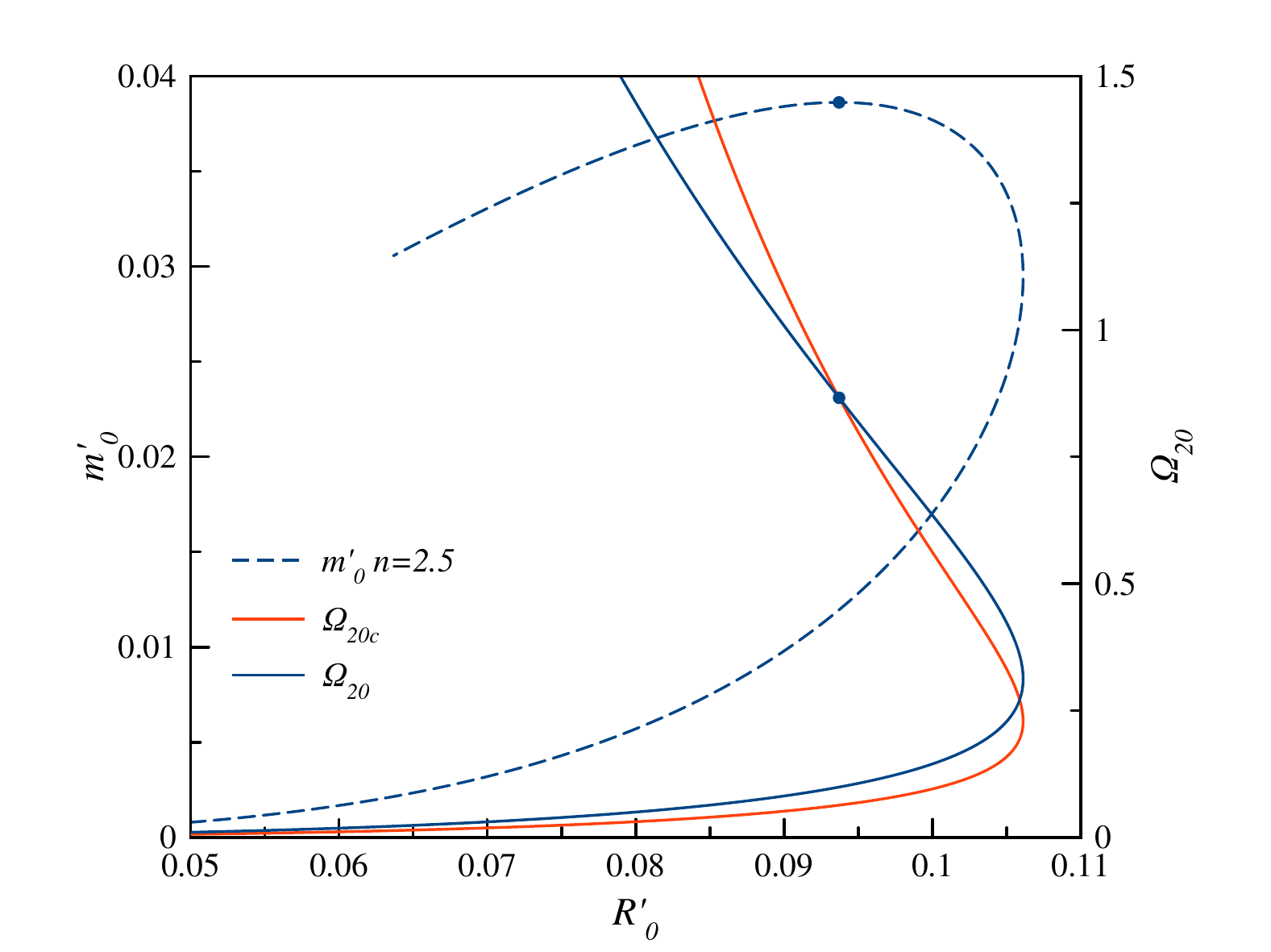}
\caption{The figure shows the $m'_0=m'_0(R'_0)$ diagram for the EOS $P=\beta \sigma^{2.5}$, as well as the curves for $\Omega_{20c}$ 
and $\Omega_{20}$.}
    \label{n2.5}
\end{figure}

\begin{figure}
    \centering
    \includegraphics[width=0.5\textwidth]{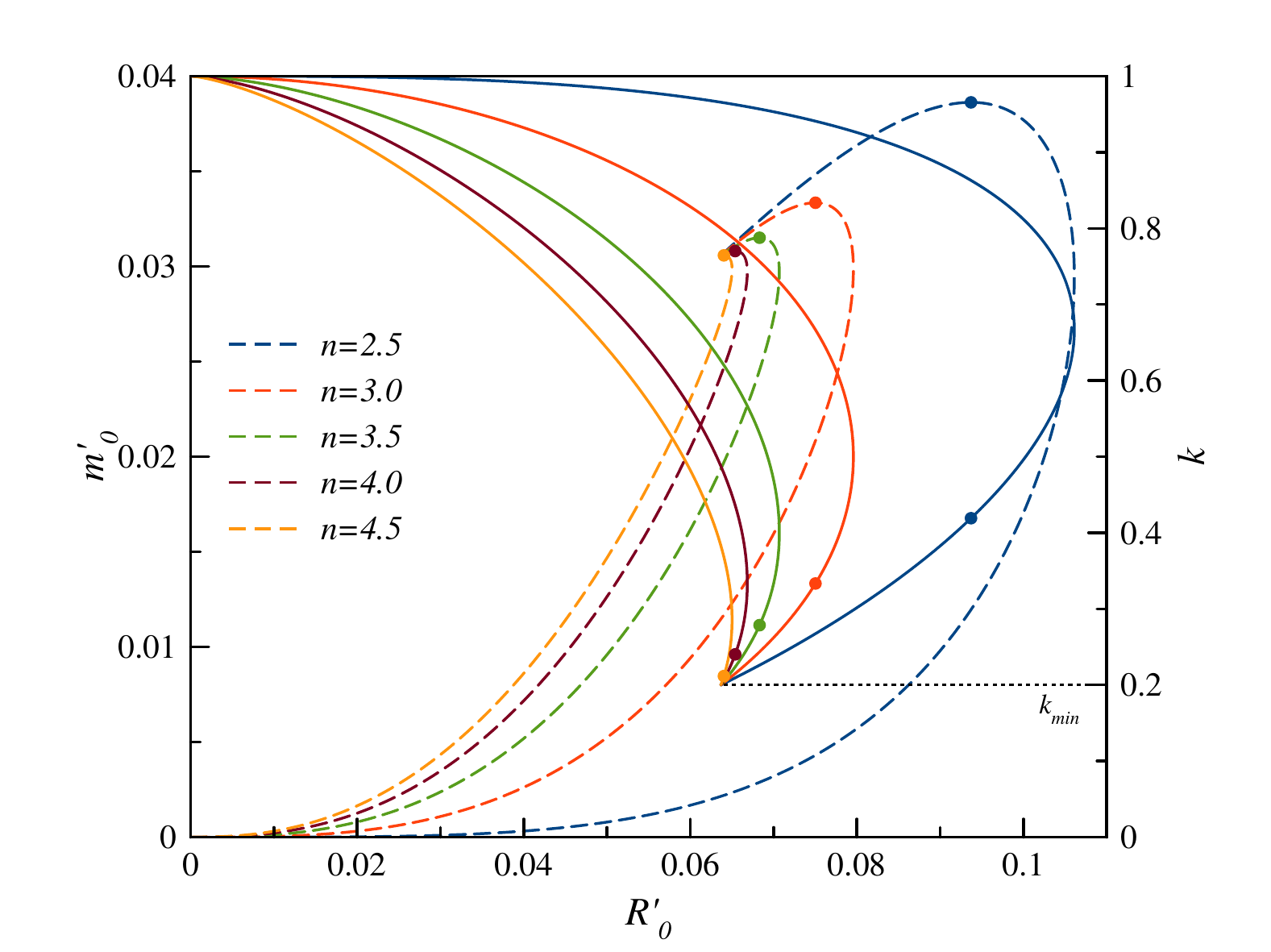}
    \includegraphics[width=0.48\textwidth]{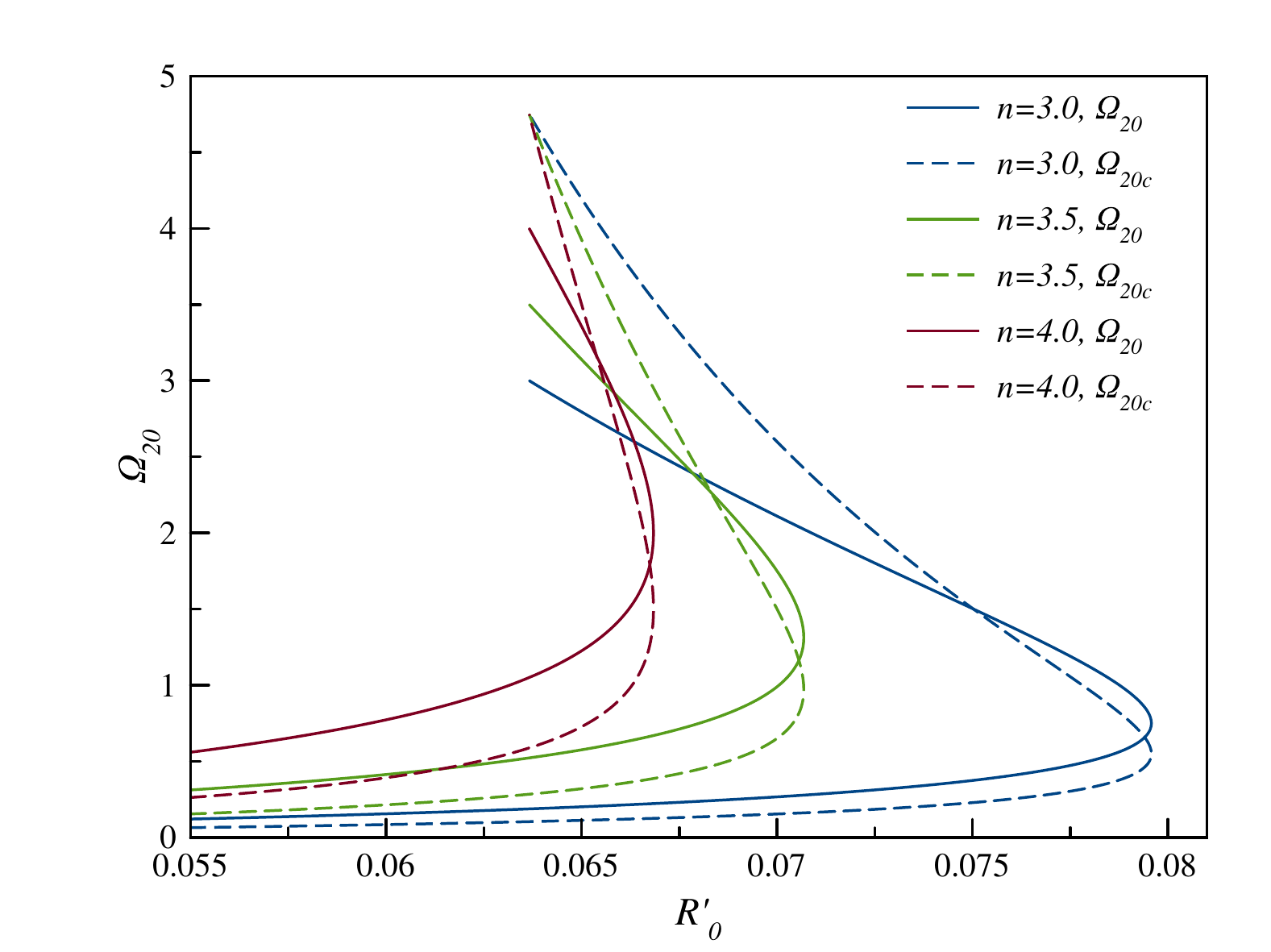}
    \caption{Upper panel: $m_0'=m_0'(R_0')$ 
    (dashed line)
    and $k=k(R_0')$ curves (full line) for EOS II and several values of $n$. $k_{min}=1/5$ is the smallest value of $k$ which satisfies the DEC. Lower panel: plots of $\Omega_{20}$ 
and $\Omega_{20c}$ in terms of $R'_0$ for different values of $n$ in EOS II.  }
    \label{figmkEOSII}
\end{figure}

Let us to study now the EOS II given by
\begin{align}
    P&=\beta\sigma^n, \label{eos3}
\end{align}
where $[\beta]=L^{n-1}$. The case $n=2$ corresponds to the  case studied in Subsection \ref{quad}. Equations (\ref{rps2}) and (\ref{mps2}) now read
\begin{align}
    R_0(\sigma_0)&=\frac{\beta\sigma_0^{n-2}}{\pi(1+4\beta\sigma_0^{n-1})}, \label{rps3}\\
    m_0(\sigma_0)&=4\pi\sigma R_0^2 (1-2\pi\sigma_0 R_0). \label{mps3}
\end{align}
Defining the dimensionless quantities $\sigma'$, $P'$, $R'$ and $m'$
by
\begin{align}
    \sigma'&=\beta^{\frac{1}{n-1}}\sigma, \label{s3l}\\
    P'&=\beta^{\frac{1}{n-1}}P,\label{p3l}\\
    R'&=\beta^{\frac{1}{1-n}}R,\label{r3l}\\
    m'&=\beta^{\frac{1}{1-n}}m,\label{m3l}
\end{align}
equations (\ref{eos3}), (\ref{rps3}) and (\ref{mps3}) read
\begin{align}
     P_0'&=\sigma_0'^n, \label{eos3l}   \\
    R_0'&=\frac{\sigma_0'^{n-2}}{\pi(1+4\sigma_0'^{n-1})}, \label{rps3l}\\
    m_0'&=4\pi\sigma_0' R_0'^2 (1-2\pi\sigma_0' R_0'). \label{mps3l}
\end{align}
Equations (\ref{rps3l}) and (\ref{mps3l}) can be solved numerically to build the $m'_0=m'_0(R'_0)$ curves, an example of which is shown in Figure
\ref{n2.5} (for $n=2.5$).
The curves corresponding to $\Omega_{20c}$ and $\Omega_{20}$ are also shown.
The stability interval goes from the maximum of the $m'_0=m'_0(R'_0)$ curve (which again coincides with the crossing of the $\Omega_{20}$ curves), all the way down to small values of $m'_0$ and $R'_0$.
The upper panel of Fig. \ref{figmkEOSII}
shows the $m'_0=m'_0(R'_0)$ curves for $n=2.5,3.0,3.5,4.0,4.5$ (dashed line), the associated $k=k(R_0)$ curves (Eq. \eqref{RN5}) for each case (full line), and the smallest value of $k$ ($k_{min}=1/5$) which satisfies the DEC (these curves will be useful below). The   
$\Omega_{20c}$ 
and $\Omega_{20}$
curves are displayed in the lower panel. In all cases,
the maximum of the 
$m_0'=m_0'(R_0')$ curve coincides with the crossing of the $\Omega_2$ curves. 
Also, 
all cases approach $m'=R'=0$ for $\sigma'\rightarrow \infty$ (high-energy limit, not shown due to the DEC violation) and  $\sigma'\rightarrow 0$ (low-energy limit), and verify the condition $2m_0'< R_0'$.
Notice also that in the dimensionless quantities we are using, all the 
$m_0'=m_0'(R_0')$ curves end at the same point, where the DEC is marginally satisfied (\emph{i.e.} $P_0'=\sigma_0'$).

\section{A non-barotropic EOS (EOS III)}
\label{sec:5}
\begin{figure}
    \centering
    \includegraphics[width=0.5\textwidth]{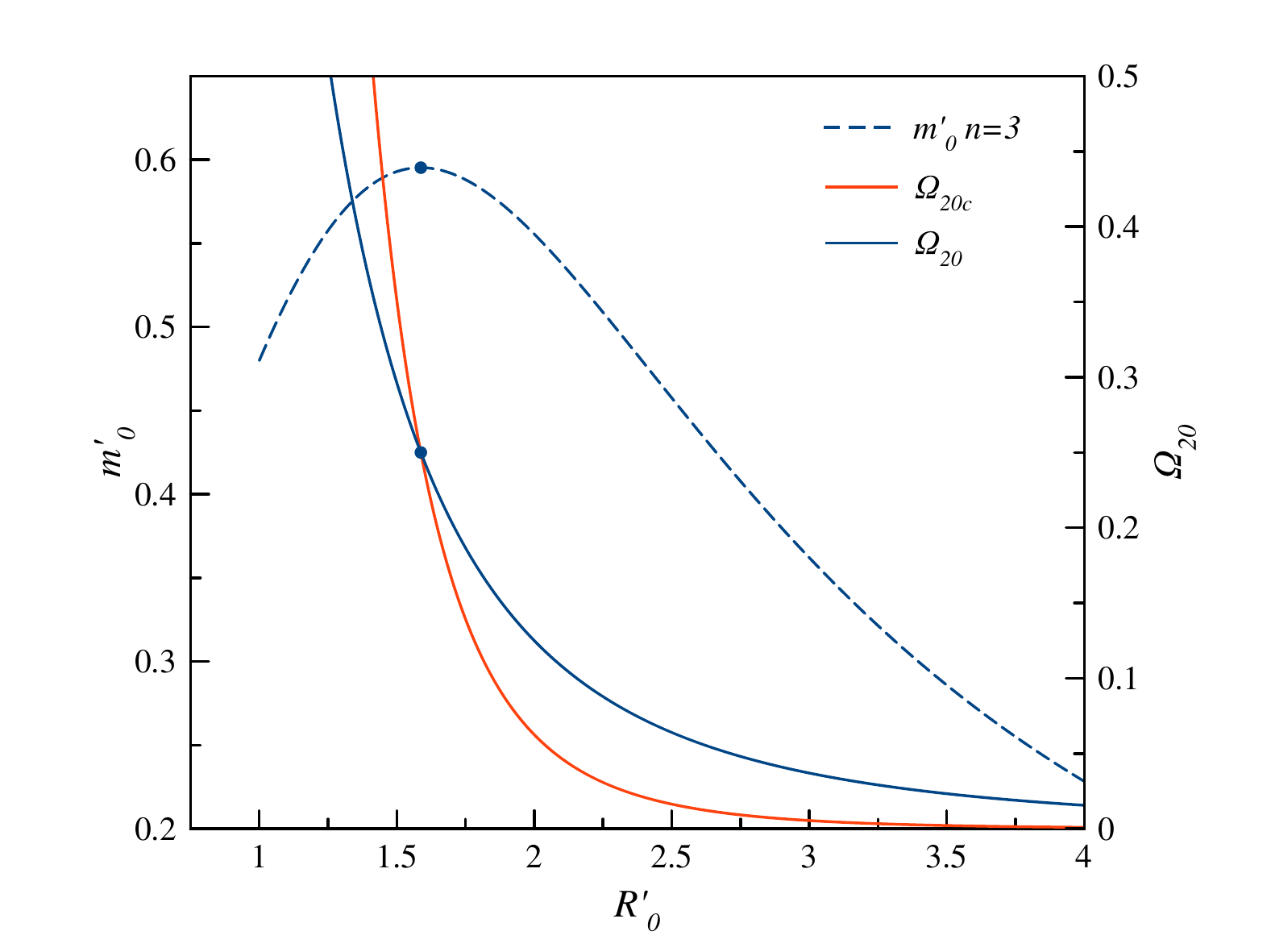}
    \caption{$m'_0=m'_0(R'_0)$ diagram, as well as $\Omega_{20}$ and 
    $\Omega_{20c}$ curves for EOS III with $n=3$.}
    \label{nbn3}
\end{figure}

\begin{figure}
    \includegraphics[width=0.5\textwidth]{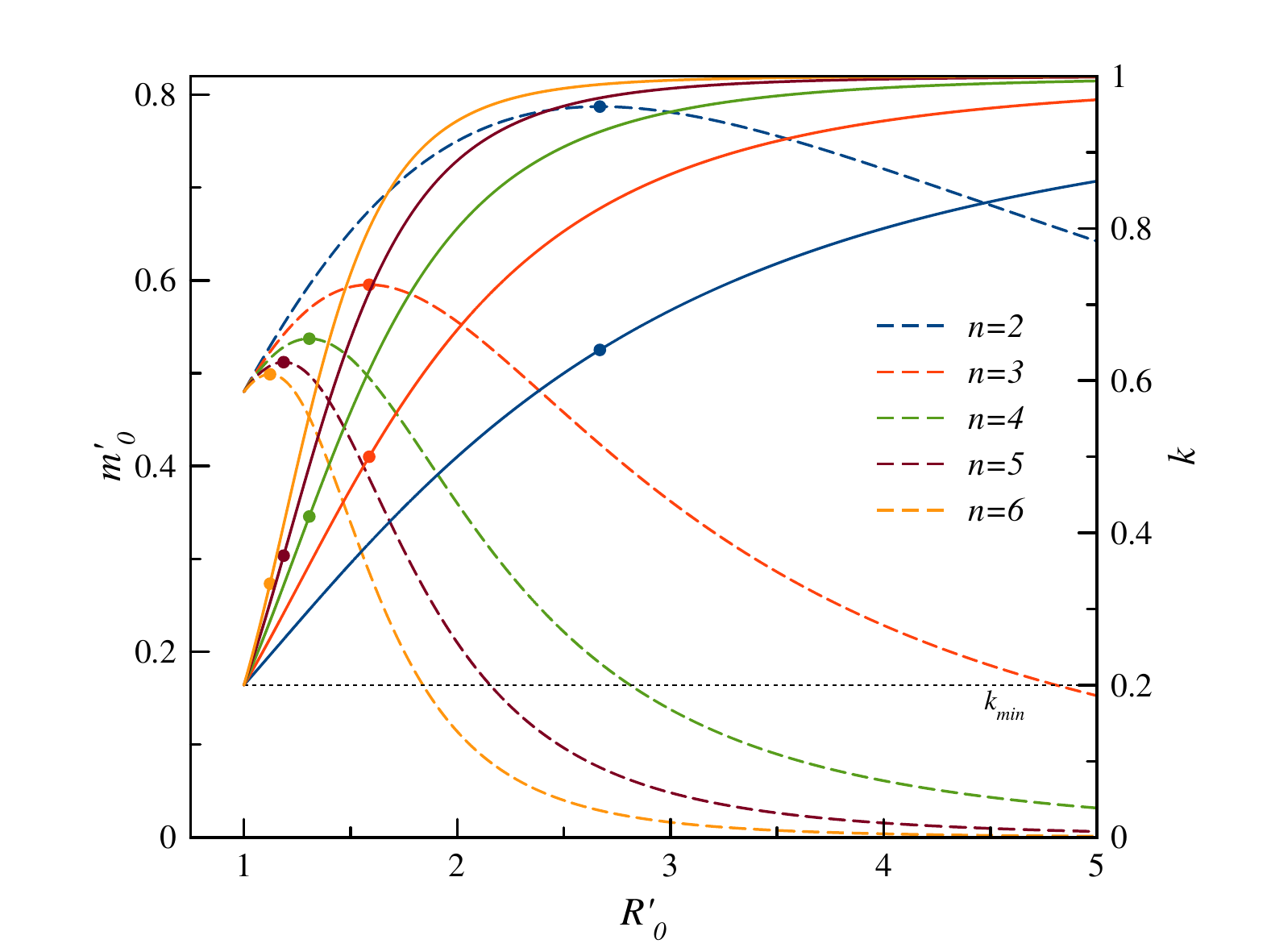}
    \includegraphics[width=0.48\textwidth]{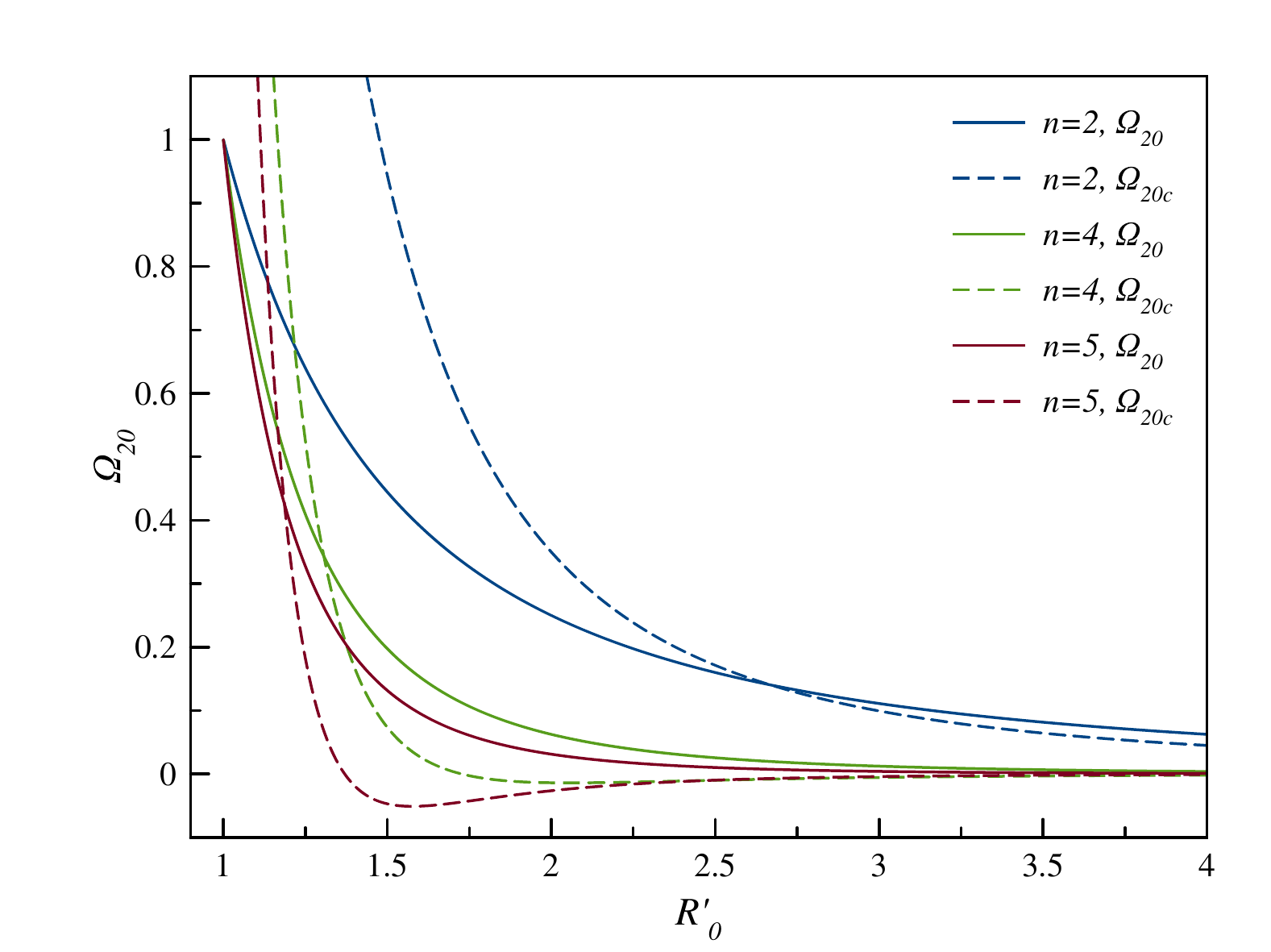}
    \caption{Upper panel: The figure shows the $m_0=m_0(R_0)$ (dashed line) and $k=k(R_0)$ curves (full line) for EOS III and several values of $n$. $k_{min}=1/5$ is the smallest value of $k$ which satisfies the DEC. Lower panel: $\Omega_{20}$ and
    $\Omega_{20c}$ curves for different values of $n$ for the non-barotropic EOS III.  }
    \label{figmkEOSIII}
\end{figure}

In this section we shall explore 
the dynamically stable equilibrium configurations that follow from the non-barotropic EOS  given by \cite{Varela2013}
\begin{align}
    P(\sigma,R)&=\frac{A}{R^n}\sigma, \label{psrnb}
\end{align}
where $[A]=L^n$.
Defining for convenience the dimensionless energy density, pressure, radius and mass as
\begin{align}
    \sigma'&= \sigma A^{1/n}, \label{slnb}\\
    P_0'&= P A^{1/n}, \label{plnb}\\
    R'&= \frac{R}{A^{1/n}}, \label{rlnb}\\
    m'&= \frac{m}{A^{1/n}}, \label{mlnb}
\end{align}
and using the EOS in the equations for the mechanical equilibrium of the shell, given in
Eqs. \eqref{rps2} and \eqref{mps2},
the following $m_0'(R_0')$ relation is obtained:
\begin{align}
    m_0'(R_0')&= \frac{4R_0'(R_0'^n+2)}{(R_0'^n+4)^2}.  \label{mxrnb}
\end{align}
The extremum is given at
\begin{align}
    R_{0m}'&=\left(\frac{3+\sqrt{8n+1}}{n-1}\right)^{1/n}\;\;\;,\; {\rm for}\; n>1. \label{rmnb}
\end{align}
The maximum in the $m_0'=m_0'(R_0')$ diagram, is given by 
\begin{align}
    m_{0max}'=m_0'(R_{0m}')=&4(n-1)\left(\frac{3+\sqrt{8n+1}}{n-1}\right)^{1/n}\nonumber
    \\
    &\left(\frac{\sqrt{8n+1}+2n+1}{\sqrt{8n+1}+4n-1}\right),\;\;n>1. \label{mmnb}
\end{align}
The $m'_0=m'_0(R'_0)$ curve
for $n=3$ is shown in Fig. \ref{nbn3}.
The stable equilibrium configurations are those located to the right of the maximum, in correlation with the crossing of the $\Omega_2$ curves. Notice that there are configurations with very small $m'_0$ and very large $R'_0$. The $m'_0=m'_0(R'_0)$ curves for several values of $n$ are shown in the upper panel of Fig. \ref{figmkEOSIII} (dashed line) together with the corresponding $k(R'_0)$ curves (full line) while the associated $\Omega_2$ curves are displayed in the lower panel. In all cases,
the maximum of the 
$m_0'=m_0'(R_0')$ curve coincides with the crossing of the $\Omega_2$ curves.

\section{Thermodynamical stability}
\label{sec:6}
We shall compare 
next the results presented in the previous sections (via dynamical stability) with those originating from thermodynamical stability. This type of stability was studied 
for the neutral shell
in \cite{Martinez1996}. Two equations of state were used there, a phenomenological one for the temperature,  and the one that follows from the junction conditions (Eq. \eqref{pMR}), to obtain the entropy of a thin shell with constant number of particles. Starting  from the first law:
\begin{equation}
    dS = \beta dM + p dA,
\label{law1}
\end{equation}
where $\beta \equiv 1/T$ and $A\equiv 4\pi R^2$,
Martinez showed that, as a consequence of Eq. \eqref{pMR} and the integrability condition for $S$, the function $\beta$ must have the form
\begin{equation}
    \beta(M,R)=b(r_+)k,
    \label{beta}
\end{equation}
where $b$ is an arbitrary function of $r_+(M,R_0)=2m(M,R_0)$, and $k$ is given in Eq. \eqref{RN5}.
The explicit form of the function $b$ should be obtained from an explicit model of the matter that composes the shell. The following phenomenological form was chosen in 
\cite{Martinez1996}:
\begin{equation}
    b(r_+;\eta,a)=\frac{2\eta}{{\ell_P}^{1+a}}\:r_+^a, \label{brplus}
\end{equation}
where $\eta$ and $a$ are dimensionless coefficients. Such a choice leads to 
\begin{equation}
    S(M,R_0;\eta,a)=\frac{\eta}{1+a}\left( \frac{r_+}{{\ell}_P}\right)^{1+a}+S_0,
\end{equation}
for $a\neq -1$. This expression reduces to the Bekenstein-Hawking entropy for $a=1$ and $\eta = 2\pi$. By demanding that a zero mass shell possess null entropy, it follows that $S_0=0$ and $a>-1$.\\
The regions of thermodynamical stability in the $(M,R)$ plane are determined by the conditions
\begin{eqnarray}
\left( \frac{\partial^2 S}{\partial M^2}\right)_A \leq 0,\\
\left( \frac{\partial^2 S}{\partial A^2}\right)_M \leq 0,\\
\left( \frac{\partial^2 S}{\partial M^2}\right)_A
\left( \frac{\partial^2 S}{\partial A^2}\right)_M
-\left( \frac{\partial^2 S}{\partial M\partial A}\right)^2
\geq 0.
\end{eqnarray}
As shown in \cite{Martinez1996}, 
these conditions are more concisely expressed in terms of $k$. Together with the normalization of the entropy and the DEC, they imply that 
\begin{align*}
-1 &< a \leq \frac{12}{19}
\end{align*}
for a stable shell. 
For $-1<a\leq 0$ 
the stability conditions do not constrain the value of $k$, hence the sole constraint for stability is that coming from the dominant energy condition, namely 
$R\geq \frac{25}{24}r_+$, or $k\geq 1/5$.
The stability condition with the crossed derivatives leads to
\begin{equation}
(3a+1)k^2+2ak+(a-1)\leq 0,
\label{crossed}
\end{equation}
which is automatically satisfied for $a\leq 0$, but restricts the values of $k$ for $0<a\leq 12/19$.

\section{Dynamical and thermodynamical stability}
\label{sec:7}
\begin{figure}
    \centering
    \includegraphics[width=0.5\textwidth]{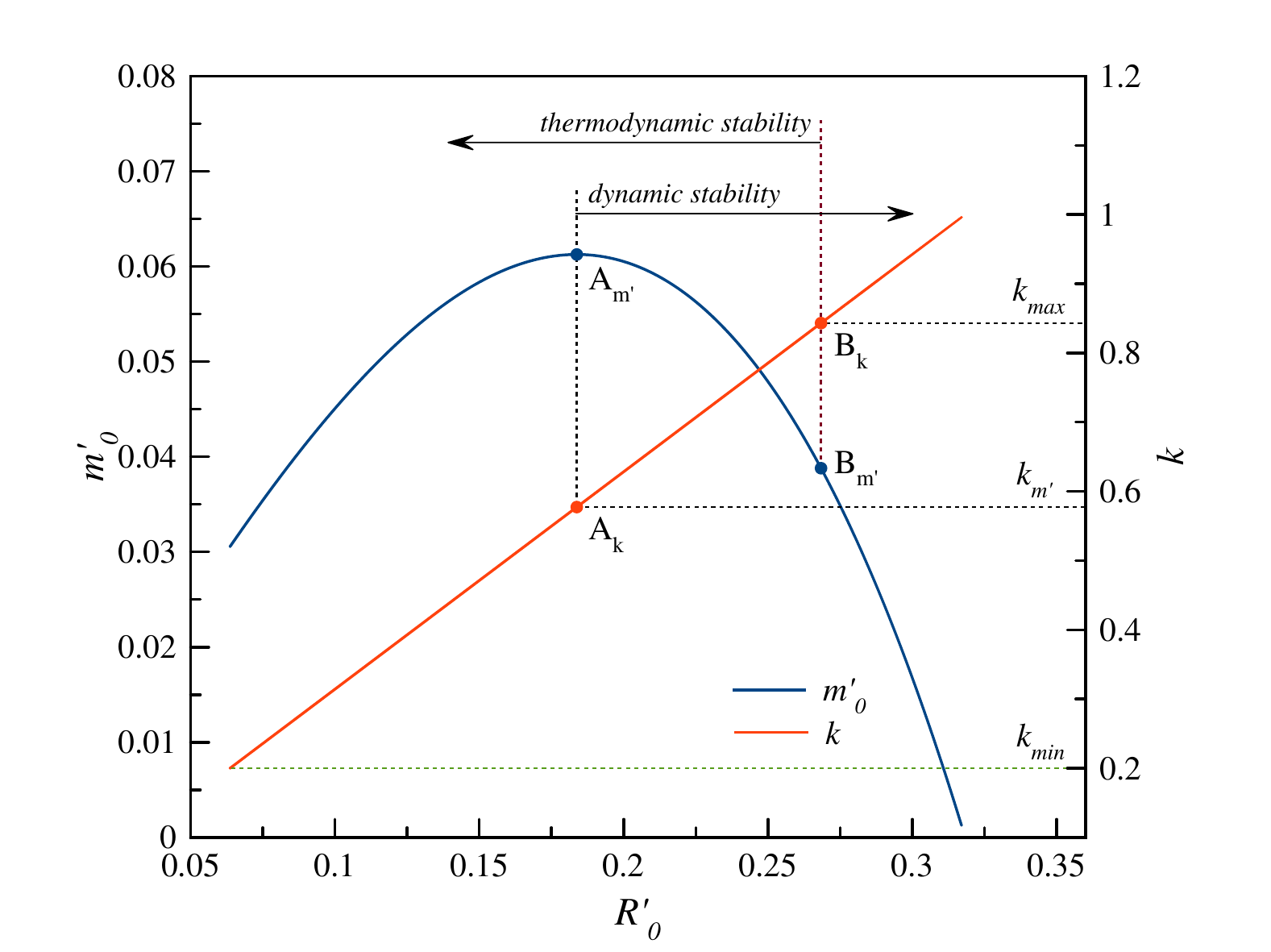}
    \caption{Dynamic and thermodynamic stability regions for EOS II with $n=2$ and $a=0.06$. The states with $k<k_{min}=1/5$ (which corresponds to $a>a_{max}=12/19$) violate the DEC. The allowed values of $k$ for thermodynamic stability satisfy $k_{min}< k <k_{max}$, 
    where $k_{max}=0.843130$ follows from Eq. (\ref{crossed}).
    The allowed values of $k$ for dynamical stability satisfy $k_{m'}<k$, where $k_{m'}=1/\sqrt{3}$ corresponds to 
    the maximum of the curve $m'_0=m'_0( R'_0)$.}
    \label{figb}
\end{figure}

\begin{figure}
    \centering
    \includegraphics[width=0.5\textwidth]{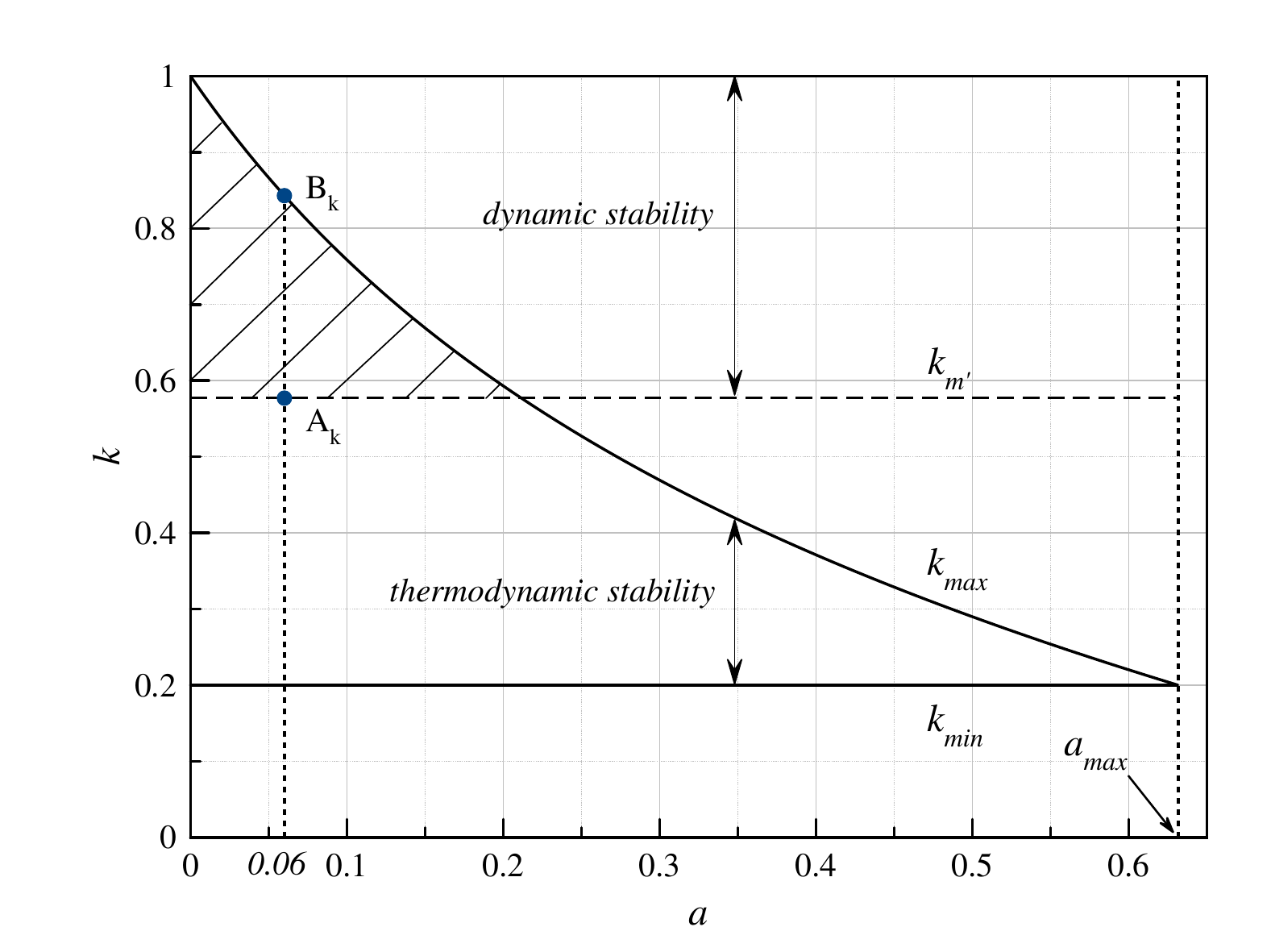}
    \caption{Dynamic and thermodynamic stability for EOS II with $n=2$ in the $(a,k)$ plane. The curve $k_{max}(a)$ follow from Eq. \eqref{crossed}. The points $A_k$ and $B_k$ are those in Fig. \ref{figb}.} 
    \label{figc}
\end{figure}

\begin{figure}
    \centering
    \includegraphics[width=0.5\textwidth]{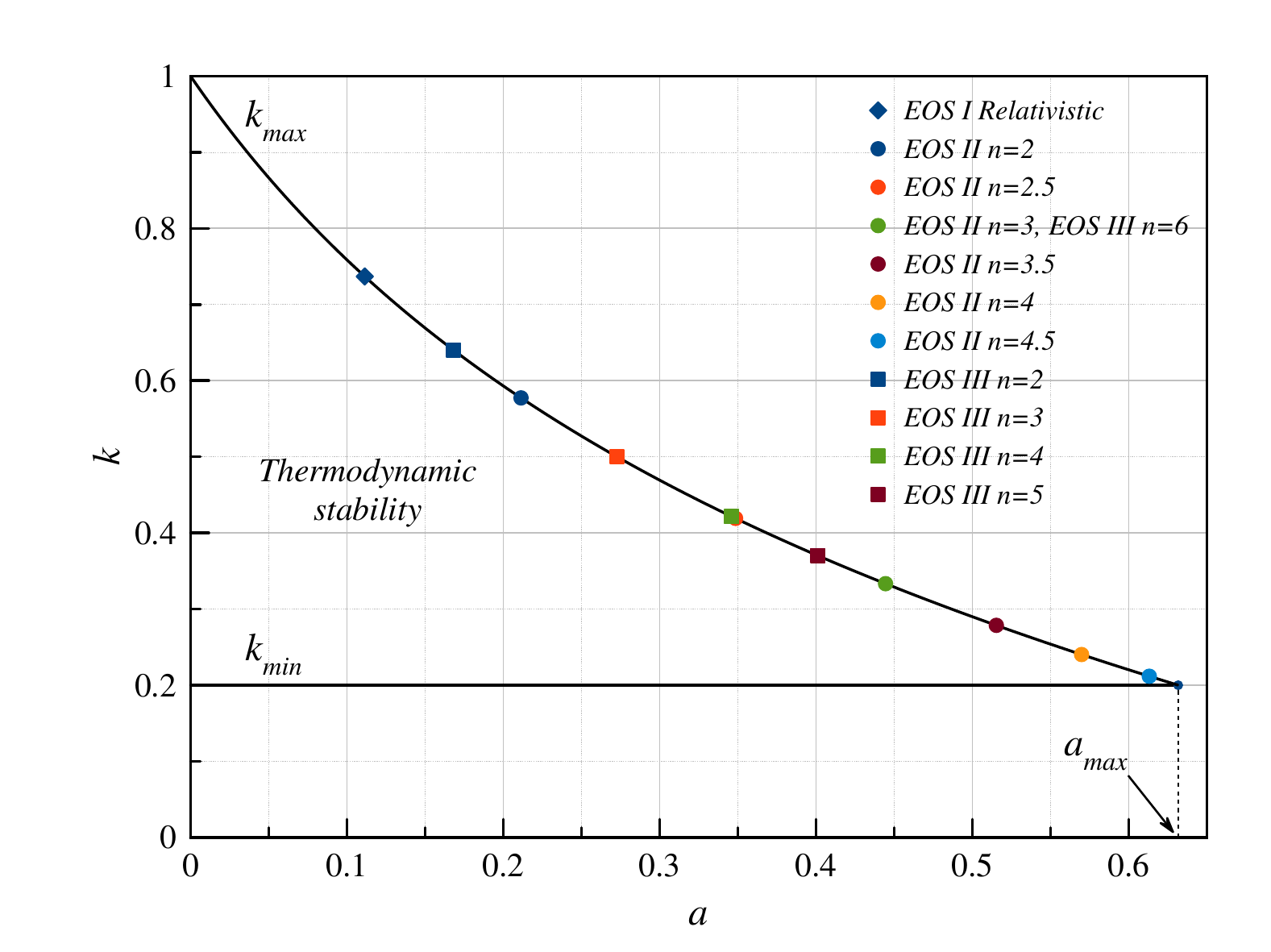}
    \caption{Dynamic and thermodynamic stability for all the EOS examined here. The states with $k<k_{min}=1/5$ (which corresponds to $a>a_{max}=12/19$) violate the DEC. The allowed values of $k$ for thermodynamic stability are those below the $k_{max}(a)$ curve, determined by Eq. \ref{crossed}, and above $k_{min}$. The allowed values of $k$ for dynamical stability are those above the value of $k$ on the curve for each EOS, which corresponds to the value of $k_{m'}$ of the maximum of the curve $m'_0=m'_0(R'_0)$ for each EOS.} 
    \label{figa}
\end{figure}

Before presenting the results for all the EOS considered here with the anzat of Eq. \eqref{brplus}, let us give an example. Figure \ref{figb} shows the $m'_0=m'_0(R'_0)$ diagram for the EOS II with $n=2$, and the values of $k$ obtained from its definition, using the equilibrium curve 
$m'_0=m'_0(R'_0)$.
For the case at hand, such a curve is given by Eq. (\ref{mlrlnr}), which results in $k(R'_0)=\pi R'_0$.
The points that are dynamically stable are those on the curve to the right 
of $A_{m'}$, which correspond to the interval $k>k_{m'}=1/\sqrt{3}$.
The thermodynamical stability region is $k_{min}<k<k_{max}$, where $k_{max}=0.843130$ is determined by Eq. (\ref{crossed}) with $a =0.06$. Hence, the 
states that are both thermodynamically and dynamically stable are those between the points $A_{m'}$ and $B_{m'}$ in the $m'_0=m'_0(R'_0)$ curve.
Figure \ref{figc} 
presents the region in the $(a,k)$ plane which correspond to configurations that are thermodynamically stable (which are those between the horizontal line $k_{min}$ and the curve $k_{max}=k_{max}(a)$), as well as that of dynamically stable configurations (which are those above the horizontal line $k_{m'}$). The intersection of these regions yields the set of points $(a,k)$ associated to configurations that are both thermodynamically and dynamically stable.
Figure \ref{figa}
presents the results in the $(a,k)$ plane for all the EOS we have examined. 
For a given value of $a$, the thermodynamically stable configurations are those with $k_{min}<k<k_{max}$. Also shown in the plot are the points at which the horizontal line associated to the value of $k$  corresponding to the maximum of the $m'_0=m'_0(R'_0)$ curve ($k_{m'}$) intersects the $k_{max}=k_{max}(a)$ curve, obtained from Eq. \eqref{crossed}. For a given EOS, the configurations that are both thermodynamically and dynamically stable are those above the horizontal line that crosses the corresponding point and below the $k_{max}=k_{max}(a)$ curve. 
The plot clearly shows that 
the requirement of thermodynamical stability greatly restricts the region of the parameter space allowed by the dynamical stability. 
It also follows from our results that the 
region associated to dynamical stability is larger for larger values of $n$, for both types of EOS considered here. This is to be expected since a larger $n$ is associated with more stiff matter.

\section{Closing remarks}
\label{sec:8}
We have examined the dynamical and thermodynamical stability of neutral mass shells for different equations of state, described by $P=P(\sigma , R)$. 
To attain this goal, the condition for dynamical stability valid for an arbitrary EOS (obtained in previous works) given in terms of the derivatives 
$\Omega_1=\frac{dP}{dR}$
and $\Omega_2=\frac{dP}{d\sigma}$
was adapted to the case of a given EOS, yielding the curve $\Omega_{20c}=\Omega_{20c}(R_0')$. The result was compared to the 
$\Omega_2=\Omega_2(R_0')$
curve that follows from the given EOS and the equations for a shell in equilibrium, to determine the dynamically stable configurations. 
Using the criteria for thermodynamical stability and a specific form for the entropy of the shell, the set of thermodynamically stable configurations was also determined. 
The main result is that thermodynamical stability greatly constraints set of stable equilibrium configurations determined by dynamical stability, due to the 
upper constraint
defined by Eq. (\ref{crossed}).
Our results also 
confirm that stable states are those to the right of the maximum of the 
$m'_0=m'_0(R'_0)$ curve. Such a link between the 
maximum mass along a sequence of equilibrium configurations of the shell and the onset of stability was obtained in \cite{LeMaitre2019} for the case of a barotropic EOS. Our findings suggest that it can be extended to EOS of the type $P=P(\sigma, R)$.\\
The generalization and consequences of our results 
to the interesting cases of
 self-gravitating shells in $d$ dimensions \cite{Andre2019}, 
and charged 
 \cite{Lemos2015}
 and rotating 
 \cite{Lemos2015c}
 shells are left for future work. 
\begin{acknowledgements}
This work was supported by PROSNI 2018-2019, Conacyt M\'exico and Universidad de Guadalajara.

\end{acknowledgements}

\section*{Appendix: 2d fermion gas}
We obtain here the EOS of a two-dimensional relativistic ideal Fermi gas of $N$ particles of mass $m$ at $T=0$ in a square of side $L$. In the thermodynamical limit we have $L\rightarrow\infty$, $N\rightarrow\infty$ and the particle number density $n=N/L^2$ constant.\\
For the summation over states $q$, we have in the continuous case
\[
\sum_q \longrightarrow \frac{2L^2}{h^2}\int d^2p,
\]
where $p$ is the two-dimensional linear momentum. Then, for the particle number density $n=N/L^2$ we have
\begin{align}
    N&=\frac{2L^2}{h^2}\int_0^{p_F} d^2p=
    \frac{L^2}{2\pi\hbar^2}p_F^2, \nonumber\\
    n&=\frac{p_F^2}{2\pi\hbar^2}=\frac{x^2}{2\pi\lambda^2},\label{n1}
\end{align}
where $p_F$ is the Fermi momentum, $x=p_F/mc$, and $\lambda=\hbar/mc$ is the Compton wavelength.\\ 
Now we obtain the internal energy density $\sigma$. For the internal energy $U$ of the gas we have
\begin{align}
    U&=\frac{2L^22\pi}{h^2}\int_0^{p_F}\sqrt{p^2c^2+m^2c^4}\,p\,dp \nonumber\\
    \sigma&=U/L^2=\frac{1}{\pi\hbar^2}\int_0^{p_F}\sqrt{p^2c^2+m^2c^4}\,p\,dp,\nonumber \\
    \sigma&=\frac{mc^2}{3\pi\lambda^2}\left[\left(x^2+1\right)^{3/2}-1\right].\label{u1}
\end{align}
For the pressure we have
\begin{align}
    P&=\frac{1}{2}\frac{1}{\pi\hbar^2}\int_0^{p_F}\frac{p^2c^2}{\sqrt{p^2c^2+m^2c^4}}\,p\,dp, \nonumber \\
    P&=\frac{mc^2}{3\pi\lambda^2}\left[\frac{1}{2}(x^2-2)\sqrt{x^2+1}+1\right]. \label{p1.5}
\end{align}
Finally, the EOS for a two-dimensional ideal relativistic Fermi gas at $T=0$ is given by
\begin{align}
   n&=\frac{x^2}{2\pi\lambda^2},\\
    \sigma&=\frac{mc^2}{3\pi\lambda^2}\left[\left(x^2+1\right)^{3/2}-1\right], \label{sigmar}\\
   P&=\frac{mc^2}{3\pi\lambda^2}\left[\frac{1}{2}(x^2-2)\sqrt{x^2+1}+1\right]. \label{pr}
\end{align}

%
%


\bibliographystyle{spphys}       

\bibliography{bibliography}

\begin{thebibliography}{10}
\providecommand{\url}[1]{{#1}}
\providecommand{\urlprefix}{URL }
\expandafter\ifx\csname urlstyle\endcsname\relax
  \providecommand{\doi}[1]{DOI \discretionary{}{}{}#1}\else
  \providecommand{\doi}{DOI \discretionary{}{}{}\begingroup
  \urlstyle{rm}\Url}\fi

\bibitem{York1986}
J.W. York, Jr., Phys. Rev. \textbf{D33}, 2092 (1986).
\newblock \doi{10.1103/PhysRevD.33.2092}

\bibitem{Lemos2015}
J.P.S. Lemos, G.M. Quinta, O.B. Zaslavski, Phys. Rev. \textbf{D91}(10), 104027
  (2015).
\newblock \doi{10.1103/PhysRevD.91.104027}

\bibitem{Lemos2015b}
J.P.S. Lemos, G.M. Quinta, O.B. Zaslavskii, Phys. Lett. \textbf{B750}, 306
  (2015).
\newblock \doi{10.1016/j.physletb.2015.08.065}

\bibitem{Lemos2016}
J.P.S. Lemos, G.M. Quinta, O.B. Zaslavskii, Phys. Rev. \textbf{D93}(8), 084008
  (2016).
\newblock \doi{10.1103/PhysRevD.93.084008}

\bibitem{Lemos2017}
J.P.S. Lemos, M.~Minamitsuji, O.B. Zaslavskii, Phys. Rev. \textbf{D96}(8),
  084068 (2017).
\newblock \doi{10.1103/PhysRevD.96.084068}

\bibitem{Lemos2007}
J.P.S. Lemos, O.B. Zaslavskii, Phys. Rev. \textbf{D76}, 084030 (2007).
\newblock \doi{10.1103/PhysRevD.76.084030}

\bibitem{Lemos_2011}
J.P. Lemos, O.B. Zaslavskii, Physics Letters B \textbf{695}(1-4), 37–40
  (2011).
\newblock \doi{10.1016/j.physletb.2010.11.033}.
\newblock \urlprefix\url{http://dx.doi.org/10.1016/j.physletb.2010.11.033}

\bibitem{Kijowski2006}
J.~Kijowski, G.~Magli, D.~Malafarina, General Relativity and Gravitation
  \textbf{38}(11), 1697 (2006).
\newblock \doi{10.1007/s10714-006-0323-0}.
\newblock \urlprefix\url{https://doi.org/10.1007/s10714-006-0323-0}

\bibitem{Martinez1996}
E.A. Martinez, Phys. Rev. \textbf{D53}, 7062 (1996).
\newblock \doi{10.1103/PhysRevD.53.7062}

\bibitem{Brady1991}
P.R. Brady, J.~Louko, E.~Poisson, Phys. Rev. D \textbf{44}, 1891 (1991).
\newblock \doi{10.1103/PhysRevD.44.1891}.
\newblock \urlprefix\url{https://link.aps.org/doi/10.1103/PhysRevD.44.1891}

\bibitem{Habib2017}
S.~Habib~Mazharimousavi, M.~Halilsoy, S.N. Hamad~Amen, Int. J. Mod. Phys.
  \textbf{D26}(14), 1750158 (2017).
\newblock \doi{10.1142/S0218271817501589}

\bibitem{Rahaman2007}
F.~Rahaman, M.~Kalam, S.~Chakraborty, Acta Phys. Polon. \textbf{B40}, 25 (2009)

\bibitem{Varela2013}
V.~Varela, Phys. Rev. \textbf{D92}, 044002 (2015).
\newblock \doi{10.1103/PhysRevD.92.044002}

\bibitem{Garcia2011}
N.M. Garcia, F.S.N. Lobo, M.~Visser, Phys. Rev. \textbf{D86}, 044026 (2012).
\newblock \doi{10.1103/PhysRevD.86.044026}

\bibitem{Kim2014}
H.C. Kim, Phys. Rev. D \textbf{89}, 064001 (2014).
\newblock \doi{10.1103/PhysRevD.89.064001}.
\newblock \urlprefix\url{https://link.aps.org/doi/10.1103/PhysRevD.89.064001}

\bibitem{MartinMoruno2011}
P.~Martin~Moruno, N.~Montelongo~Garcia, F.S. Lobo, M.~Visser, JCAP \textbf{03},
  034 (2012).
\newblock \doi{10.1088/1475-7516/2012/03/034}

\bibitem{Guo2005}
Z.K. Guo, Y.Z. Zhang, Phys. Lett. \textbf{B645}, 326 (2007).
\newblock \doi{10.1016/j.physletb.2006.12.063}

\bibitem{Debnath2007}
U.~Debnath, Astrophys. Space Sci. \textbf{312}, 295 (2007).
\newblock \doi{10.1007/s10509-007-9690-6}

\bibitem{LeMaitre2019}
P.~LeMaitre, E.~Poisson, Am. J. Phys. \textbf{87}(12), 961 (2019).
\newblock \doi{10.1119/10.0000026}

\bibitem{Israel1966}
W.~Israel, Il Nuovo Cimento B Series 10 \textbf{44}(1), 1 (1966).
\newblock \doi{10.1007/bf02710419}.
\newblock \urlprefix\url{https://doi.org/10.1007/bf02710419}

\bibitem{Andre2019}
R.~André, J.P. Lemos, G.M. Quinta, Phys.\ Rev.\ D \textbf{99}(12), 125013
  (2019).
\newblock \doi{10.1103/PhysRevD.99.125013}

\bibitem{Lemos2015c}
J.P.S. Lemos, F.J. Lopes, M.~Minamitsuji, J.V. Rocha, Phys.\ Rev.\ D
  \textbf{92}(6), 064012 (2015).
\newblock \doi{10.1103/PhysRevD.92.064012}

\end{thebibliography}
%
%

\end{document}